\documentclass[twocolumn,english,prl]{revtex4-1}
\usepackage[T1]{fontenc}
\usepackage[latin9]{inputenc}
\usepackage{color}
\usepackage{amsmath}
\usepackage{amssymb}
\usepackage{graphicx}

\makeatletter
\usepackage{braket}

\makeatother

\usepackage{babel}
\begin{document}

\title{Local-in-time error in variational quantum dynamics}

\author{Rocco Martinazzo$^{1,2,*}$, Irene Burghardt$^{3}$}

\affiliation{$^{1}$Department of Chemistry, Università degli Studi di Milano,
Via Golgi 19, 20133 Milano, Italy}
\email{rocco.martinazzo@unimi.it}

\selectlanguage{english}%

\affiliation{$^{2}$Istituto di Scienze e Tecnologie Molecolari, CNR, via Golgi
19, 20133 Milano, Italy}

\affiliation{$^{3}$Institute of Physical and Theoretical Chemistry, Goethe University
Frankfurt, Max-von-Laue-Str. 7, D-60438 Frankfurt/Main, Germany}
\begin{abstract}
The McLachlan ``minimum-distance'' principle for optimizing approximate
solutions of the time-dependent Schr\"{o}dinger equation is revisited,
with a focus on the local-in-time error accompanying the variational
solutions. Simple, exact expressions are provided for this error,
which are then evaluated in illustrative cases, notably the widely
used mean-field approach and the adiabatic quantum molecular dynamics.
These findings pave the way for the rigorous development of adaptive
schemes that re-size \emph{on-the-fly} the underlying variational
manifold and thus optimize the overall computational cost of a quantum
dynamical simulation. 
\end{abstract}
\maketitle
\textbf{\emph{Introduction}}. Variational principles play a major
role in quantum dynamics since they allow to devise general strategies
to evolve wavefunctions on parametrized manifolds, in such a way to
mimic as much as possible the exact quantum mechanical evolution.
There exist at least three different time-dependent variational principles,
namely the McLachlan\cite{McLachlan1964} variational principle (MVP),
the Time-Dependent Variational Principle\cite{Kramer1981} (TDVP)
and the Dirac-Frenkel\cite{Dirac1930,Frenkel} variational principle
(DFVP), which are known to be equivalent to each other under mild
conditions\cite{Broeckhove1988}, usually satisfied in practice. However,
these three variational principles have different origins and limitations
and, indeed, only the first one represents a well-founded, general
optimization scheme. The reason is that the DFVP 
\begin{equation}
\braket{\delta\Psi|(i\hbar\partial_{t}-H)|\Psi}=0\label{eq:DF condition}
\end{equation}
is \emph{not}, strictly speaking, a variational principle, since it
is not a functional variation \-- in the sense that it does not refer
to an action functional \-- but just a condition which defines an
optimization problem. It closely resembles, but is stronger than,
the condition
\begin{equation}
\Re\braket{\delta\Psi|(i\hbar\partial_{t}-H)|\Psi}=0\label{eq:TDVP condition}
\end{equation}
that results from the TDVP, which is indeed a stationary-action principle,
$\delta S=\delta\int_{t_{i}}^{t_{f}}L[\Psi_{t}]dt=0$, with the real
Lagrangian (here for normalized wavefunctions)
\[
L[\Psi_{t}]=\frac{i\hbar}{2}\left(\braket{\Psi_{t}|\dot{\Psi}_{t}}-\braket{\dot{\Psi}_{t}|\Psi_{t}}\right)-\braket{\Psi_{t}|H|\Psi_{t}}
\]
This is rather appealing because of its formal resemblance with the
\emph{classical} stationary-action principle (and the ensuing possibility
of a Hamiltonian dynamics of the variational parameters\cite{Kramer1981})
but it seems flawed due to the double ended boundary condition $\ket{\delta\Psi_{t_{f}}}=\ket{\delta\Psi_{t_{i}}}=0$
which is incongruous with a \emph{first} order equation in time (the
time-dependent Schr\"{o}dinger equation) which it is meant to replace
(see \emph{e.g.} Ref. \cite{Vignale2008}). A similar stationarity
condition, 
\begin{equation}
\Im\braket{\delta\dot{\Psi}|(i\hbar\partial_{t}-H)|\Psi}=0\label{eq:MVP condition}
\end{equation}
defines the MVP which, contrary to the above two, is firmly rooted
in purely geometrical ideas. Despite this, McLachlan's principle is
perhaps the least popular of the three, firstly because the presence
of the time-derivative of the wavefunction variation ($\delta\dot{\Psi}$)
makes it less intuitive, and secondly, because the above mentioned
equivalence of the three principles led researchers to focus on the
DFVP and the TDVP which admit an immediate physical interpretation.
In this Letter we revisit the MVP ``geometrical'' principle and
exploit some basic, hitherto unexplored, consequences. Specifically,
we will consider the local-in-time error associated with the MVP and
consider its implications for variational propagation schemes.

\textbf{\emph{The McLachlan minimum-distance principle.}} Let us first
introduce some notation. In the following it is assumed that the wavefunctions
we deal with lie on a manifold $\mathcal{M}\subseteq\mathcal{H}$
(the ``variational manifold'') that admits a smooth parametrization,
\emph{i.e.,} $\ket{\Psi}\equiv\ket{\Psi(\mathbf{x})}$ where $\mathbf{x}\in\Omega\subseteq\mathbb{R}^{n}$
and $\partial\ket{\Psi}/\partial x_{i}$'s, $\partial^{2}\ket{\Psi}/\partial x_{i}\partial x_{j}$'s
are well-defined vectors of the Hilbert space $\mathcal{H}$ of the
system. For simplicity, we assume that $\mathcal{M}$ contains its
rays, in order to allow normalization of the wavefunction. The directional
derivative along $\mathbf{u}\in\mathbb{R}^{n}$ in $\mathbf{x}_{0}$
is given by
\[
\ket{\delta_{\mathbf{u}}\Psi_{0}}=\left.\frac{d\ket{\Psi(\mathbf{x}_{0}+s\mathbf{u})}}{ds}\right|_{s=0}=\sum_{i=1}^{n}u_{i}\left.\frac{\partial\ket{\Psi(\mathbf{x})}}{\partial x_{i}}\right|_{_{\mathbf{x}=\mathbf{x}_{0}}}
\]
and defines a generic ``variation'' of $\ket{\Psi_{0}}=\ket{\Psi(\mathbf{x}_{0})}$
(\emph{i.e.}, along $\mathbf{u}$). The vectors $\ket{\delta_{i}\Psi_{0}}\equiv\partial\ket{\Psi}/\partial x_{i}|_{\mathbf{x}=\mathbf{x}_{0}}$
($i=1,..n$) span a linear space of dimension $n$, denoted as $\text{T}_{0}\mathcal{M}$,
which is the space tangent to $\mathcal{M}$ in $\ket{\Psi_{0}}$.
This linear space is \emph{real, }as long as the manifold coordinates
are real parameters, which is the most general case. Occasionally,
one may make use of complex (analytic) parametrizations, and in that
case $\text{T}_{0}\mathcal{M}$ becomes a \emph{complex} linear space,
a sufficient condition for the equivalence of the above variational
principles\cite{Broeckhove1988}. More generally, we say that the
variation $\ket{\delta\Psi_{0}}\in\text{T}_{0}\mathcal{M}$ is \emph{complex
}whenever the vector $i\ket{\delta\Psi_{0}}$ is a permitted variation,
too\footnote{This condition is satisfied by \emph{any} variation when $\text{T}_{0}\mathcal{M}$
happens to be complex-linear. The converse is also true, that is if
$i\times\text{T}_{0}\mathcal{M}=\text{T}_{0}\mathcal{M}$ then $\text{T}_{0}\mathcal{M}$
is complex-linear.}, \emph{i.e.}, $i\ket{\delta\Psi_{0}}\in\text{T}_{0}\mathcal{M}$. 

Suppose we are given $\ket{\Psi_{0}}\in\mathcal{M}$ as an initial
state for a short-time dynamics of time $dt$. The best choice for
$\ket{\Psi_{0}(dt)}\in\mathcal{M}$, the time-evolved state, should
minimize the error, that is the distance from the exact solution $\ket{\Psi_{0}^{\text{exact}}(dt)}$,
$\varepsilon dt=||\Psi_{0}(dt)-\Psi_{0}^{\text{exact }}(dt)||$ (here
written in terms of error per unit time $\varepsilon$) or, equivalently,
\[
\hbar\varepsilon=||i\hbar\dot{\Psi}_{0}-H\Psi_{0}||
\]
Stationarity with respect to variations of $\ket{\dot{\Psi}_{0}}$
gives the McLachlan condition, Eq. \ref{eq:MVP condition}, for $\ket{\Psi_{0}}$
\begin{equation}
\Im\braket{\delta\dot{\Psi}_{0}|\left(i\hbar\partial_{t}-H\right)|\Psi_{0}}=0\label{eq:McLaclhan condition}
\end{equation}
where $\ket{\delta\dot{\Psi}_{0}}$ can be thought of as a limiting
difference between the tangent vectors of two neighboring paths.
The invariance under scalar multiplication directly leads to norm
conservation, since for $\ket{\delta\dot{\Psi}_{0}}=\delta\dot{\lambda}\ket{\Psi_{0}}$
(with $\delta\dot{\lambda}$ arbitrary complex) it gives
\begin{equation}
i\hbar\braket{\Psi_{0}|\dot{\Psi}_{0}}=\braket{\Psi_{0}|H|\Psi_{0}}\label{eq:gauge-condition}
\end{equation}
which implies $2\Re\braket{\Psi_{0}|\dot{\Psi}_{0}}=d\braket{\Psi_{0}|\Psi_{0}}/dt=0$.
At the same time, the \emph{gauge} is fixed to $\hbar\Im\braket{\Psi_{0}|\dot{\Psi}_{0}}=-\braket{\Psi_{0}|H|\Psi_{0}}$,
that is, precisely that of the exact solution, $i\hbar\ket{\dot{\Psi}_{0}^{\text{exact }}}=H\ket{\Psi_{0}}$.
The same conclusions follow by taking $\mathcal{M}$ a manifold of
\emph{normalized} wavefunctions, but with a free phase factor that
is then optimized\footnote{It is worth emphasizing that, under such circumstances, the above
defined ``differential'' distance depends on both the manifold (\emph{i.e.,}
the shape of the trial wavefunction)\emph{ and} the \emph{gauge}.
This becomes obvious when considering the \emph{gauge} transformation
$\ket{\Psi_{t}}=e^{i\Theta t}\ket{\bar{\Psi}_{t}}$ ($\text{where }\ket{\bar{\Psi}_{t=0}}=\ket{\Psi_{0}}$)
and computing the time-derivative at $t=0$, $\ket{\dot{\Psi}_{0}}=i\Theta\ket{\Psi_{0}}+\ket{\dot{\bar{\Psi}}_{0}}$.
Optimization of the \emph{gauge} can be achieved by considering the
appropriate variation $\ket{\delta\dot{\Psi}_{0}}=i\delta\Theta\ket{\Psi_{0}}$
in Eq. \ref{eq:McLaclhan condition}, and results in the condition
$-\hbar\Im\braket{\Psi_{0}|\dot{\Psi}_{0}}=\braket{\Psi_{0}|H|\Psi_{0}}$
that can be combined with norm conservation to give Eq. \ref{eq:gauge-condition}.}.

Next, we consider the optimization of the path. When the time-dependence
in $\ket{\Psi_{t}}$ comes \emph{only} from variational parameters,
$\ket{\delta\dot{\Psi}_{0}}$ is nothing else that an arbitrary element
of $\text{T}_{0}\mathcal{M}$. In other words, in this case holds\emph{
}
\begin{equation}
\Im\braket{\delta\Psi_{0}|\left(i\hbar\partial_{t}-H\right)|\Psi_{0}}=0\label{eq:imaginary DF condition}
\end{equation}
since $\text{T}_{0}\mathcal{M}$ is a linear space and its elements
are just the wavefunction variations. Eq. \ref{eq:imaginary DF condition}
is only apparently similar to Eq. \ref{eq:TDVP condition} (though
they both reduce to the Dirac-Frenkel condition, Eq. \ref{eq:DF condition},
for complex variations). This becomes clear when evaluating it for
$\ket{\delta\Psi_{0}}=\ket{\dot{\Psi}_{0}}$, the time derivative
of the variational solution which is a legitimate element of $\text{T}_{0}\mathcal{M}$,
since Eq. \ref{eq:imaginary DF condition} gives
\begin{equation}
\hbar\braket{\dot{\Psi}_{0}|\dot{\Psi}_{0}}=\Im\braket{\dot{\Psi}_{0}|H|\Psi_{0}}\label{eq:novel condition}
\end{equation}
which is a genuine consequence of the McLachlan principle. The same
manipulation in the TDVP gives a different (though rather important)
condition, namely energy conservation, $\Re\braket{\dot{\Psi}_{0}|H|\Psi_{0}}=\frac{1}{2}\frac{d}{dt}\braket{\Psi_{0}|H|\Psi_{0}}=0$.
Eq. \ref{eq:novel condition} gives immediately a ``boundedness theorem''
\begin{equation}
\hbar||\dot{\Psi}_{0}||\leq||H\Psi_{0}||\label{eq:boundedness theorem}
\end{equation}
but it is actually more powerful, as is shown in the following. 

\textbf{\emph{Local-in-time error}}\emph{.} The value of the distance
at the variational minimum, denoted as $\varepsilon_{\mathcal{M}}$,
\[
\varepsilon_{\mathcal{M}}[\Psi_{0}]=\hbar^{-1}\text{min}_{u\in\text{T}_{0}\mathcal{M}}||i\hbar u-H\Psi_{0}||
\]
is a functional of $\ket{\Psi_{0}}$, depending on the chosen manifold
$\mathcal{M}$. It represents the distance of the manifold\textbf{
$\mathcal{M}$ }from the exact solution in\textbf{ }$\ket{\Psi_{0}}$\textbf{,}
\emph{i.e.,} a local-in-time measure of the performance of the variational
method associated to $\mathcal{M}$. Figuratively, it gives a ``skin''
of finite thickness to the manifold $\mathcal{M}$ that locally measures
the accuracy of the variational method associated to $\mathcal{M}$,
for the given dynamical problem. Importantly, it also sets an \emph{a
posteriori} upper bound to the wavefunction error\cite{LubichBook}
\begin{equation}
||\Psi_{0}(t)-\Psi_{0}^{\text{exact}}(t)||\leq\int_{0}^{t}\varepsilon_{\mathcal{M}}[\Psi_{0}(\tau)]d\tau\label{eq:a posteriori error bound}
\end{equation}
and can thus be used confidently to minimize the error over time when
acting on $\mathcal{M}$ (see Supplemental Material, SM). Using Eq.
\ref{eq:novel condition} one easily finds
\begin{equation}
\varepsilon_{\mathcal{M}}^{2}[\Psi_{0}]=\frac{1}{\hbar^{2}}\left(||H\Psi_{0}||^{2}-\hbar^{2}||\dot{\Psi}_{0}||^{2}\right)\label{eq:DF error}
\end{equation}
which is a simple, exact expression for the local-in-time error. When
$\text{T}_{0}\mathcal{M}$ is complex-linear, this is a simple consequence
of the fact that the variational condition can be recast as an orthogonal
projection\cite{LubichBook}, namely $i\hbar\ket{\dot{\Psi}_{0}}=\mathcal{P}_{0}H\ket{\Psi_{0}}$
where $\mathcal{P}_{0}$ is the projector onto $\text{T}_{0}\mathcal{M}$;
however, this condition is not necessary for Eq. \ref{eq:DF error}
to hold, when the MVP is used. In the following, we show how $\varepsilon_{\mathcal{M}}^{2}$
can be used in practice to assess quantitatively the quality of a
variational approximation and how to improve it when necessary. 

\begin{figure}
\begin{centering}
\includegraphics[width=0.8\columnwidth]{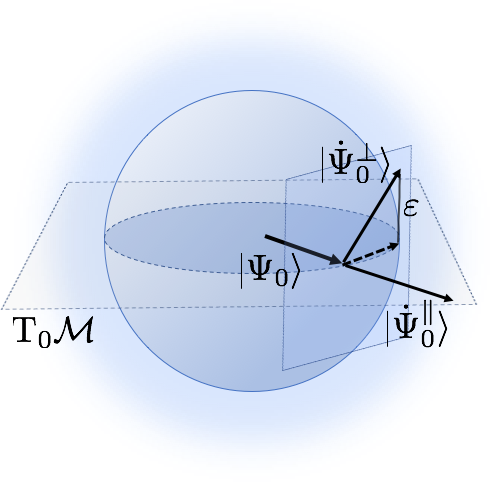}
\par\end{centering}
\caption{\label{fig:Schematics}Schematics illustrating the local-in-time error
$\varepsilon$ when $\mathcal{M}$ contains its rays and T$_{0}\mathcal{M}$
is complex-linear. Here, the sphere represents the unit sphere of
normalized vectors, and $\ket{\dot{\Psi}_{0}^{\parallel}}$ and $\ket{\dot{\Psi}_{0}^{\perp}}\equiv\ket{\dot{\Psi}_{0}^{+,\text{exact }}}$
are the ``irrelevant'' and ``relevant'' components of the exact
time derivative, given by $(i\hbar)^{-1}\bar{E}_{0}\ket{\Psi_{0}}$
and $(i\hbar\Delta E_{0})^{-1}(H-\bar{E}_{0})\ket{\Psi_{0}}$, respectively.
Note that the graphics cannot describe the fact that both components
preserve the norm. }
\end{figure}
We first rewrite Eq. \ref{eq:DF error} in a more appealing form,
since it is invariant under a shift of the Hamiltonian ($H\rightarrow H^{\epsilon}=H-\epsilon$)
provided, of course, the \emph{gauge} is modified accordingly ($\ket{\Psi_{0}}\rightarrow\ket{\Psi_{0}^{\epsilon}}=\exp(+\frac{i}{\hbar}\epsilon t)\ket{\Psi_{0}}$).
Hence, it is convenient to choose as reference energy the average
energy of the state $\ket{\Psi_{0}}$, denoted here and in the following
as $\bar{E}_{0}$, resulting in the corresponding ``standard'' \emph{gauge}
$\ket{\Psi_{0}^{\text{+}}}:=\ket{\Psi_{0}^{\bar{E}_{0}}}$. With this
\emph{gauge}, Eq. \ref{eq:DF error} takes the form 
\begin{equation}
\varepsilon_{\mathcal{M}}^{2}[\Psi_{0}]=\frac{1}{\hbar^{2}}\left(\Delta E_{0}^{2}-\hbar^{2}||\dot{\Psi}_{0}^{+}||^{2}\right)\label{eq:DF error nice}
\end{equation}
where $\Delta E_{0}^{2}=\braket{(H-\bar{E}_{0})^{2}}_{0}$ is the
energy variance and $\ket{\dot{\Psi}_{0}^{+}}$ satisfies $\braket{\Psi_{0}|\dot{\Psi}_{0}^{+}}=0$.
Again, this admits a simple interpretation since the action of $H$
on a given vector $\ket{\Psi_{0}}$ can always be split into a component
along $\ket{\Psi_{0}}$ and one orthogonal to it, $\ket{\Psi_{0}^{\perp}}$,
namely\cite{Pollak2019} 
\begin{align*}
H\ket{\Psi_{0}} & =\bar{E}_{0}\ket{\Psi_{0}}+\Delta E_{0}\ket{\Psi_{0}^{\perp}}\\
 & =i\hbar\ket{\dot{\Psi}_{0}^{\parallel}}+i\hbar\ket{\dot{\Psi}_{0}^{\perp}}
\end{align*}
where $\ket{\Psi_{0}^{\perp}}=(H-\bar{E}_{0})\ket{\Psi_{0}}/\Delta E_{0}$
is a normalized vector orthogonal to $\ket{\Psi_{0}}$. The two components\textcolor{red}{{}
}$\ket{\dot{\Psi}_{0}^{\parallel}}$ and $\ket{\dot{\Psi}_{0}^{\perp}}$
are, respectively, the ``irrelevant'' and ``relevant'' components
of the exact time-derivative (see Fig. \ref{fig:Schematics}). The
latter reduces to the time-derivative of the exact wavefunction in
the standard \emph{gauge,} $i\hbar\ket{\dot{\Psi}_{0}^{+,\text{exact }}}=i\hbar\ket{\dot{\Psi}_{0}^{\perp}}=\Delta E_{0}\ket{\Psi_{0}^{\perp}}$,\emph{
}and thus $\Delta E_{0}$ determines the ``intrinsic''\emph{ }length
of this derivative\emph{.} We note that the decomposition of Eq. \ref{eq:DF error nice}
is different from the approach of Ref. \cite{LubichBook} where the
error is written in terms of the deviation of the tangent space projection
from the exact solution. 

Interestingly, when the equations of motion can be recast in the form
$i\hbar\ket{\dot{\Psi}_{0}^{+}}=H_{v}\ket{\Psi_{0}}$, where $H_{v}$
is a ``variational'' (self-adjoint) Hamiltonian operator, the error
becomes a measure of the ability of $\mathcal{M}$ to account for
the energy fluctuations, 
\[
\varepsilon_{\mathcal{M}}^{2}[\Psi_{0}]=\frac{1}{\hbar^{2}}\left(\Delta E_{0}^{2}-\Delta E_{v,0}^{2}\right)
\]
where $\Delta E_{v,0}^{2}=\braket{\Psi_{0}|H_{v}^{2}|\Psi_{0}}$ is
the variance of the ``effective'' energy\footnote{This happens, for instance, when $\text{T}_{0}\mathcal{M}$ is complex
linear and $\mathcal{M}$ contains its rays. In that case, $H_{v}\equiv\mathcal{P}_{0}(H-\bar{E}_{0})\mathcal{P}_{0}$
and the error can also be written in the form $\hbar^{2}\varepsilon_{\mathcal{M}}^{2}[\Psi_{0}]=\braket{\Psi_{0}|(H-\bar{E}_{0})\mathcal{Q}_{0}(H-\bar{E}_{0})|\Psi_{0}}$
where $\mathcal{Q}_{0}=1-\mathcal{P}_{0}$. }. This variational energy variance is bounded, $\Delta E_{v,0}^{2}\leq\Delta E_{0}^{2}$,
and attains its maximum value for the exact solution. 

Now, upon factoring out $\Delta E_{0}^{2}$, which is common to any
manifold containing $\ket{\Psi_{0}}$, we write
\[
\varepsilon_{\mathcal{M}}^{2}[\Psi_{0}]=\frac{\Delta E_{0}^{2}}{\hbar^{2}}\left(1-r_{\mathcal{M}}^{2}[\Psi_{0}]\right)\ \ \text{with}\ \ r_{\mathcal{M}}^{2}[\Psi_{0}]:=\frac{\hbar^{2}||\dot{\Psi}_{0}^{+}||^{2}}{\Delta E_{0}^{2}}
\]
where we have introduced the dimensionless index $r_{\mathcal{M}}[\Psi_{0}]\in[0,1]$
(see Eq. \ref{eq:boundedness theorem}) with the properties
\[
\mathcal{M}'\supseteq\text{\ensuremath{\mathcal{M}}}\implies r_{\mathcal{M}'}[\Psi_{0}]\geq r_{\mathcal{M}}[\Psi_{0}]
\]
\[
r_{\mathcal{M}}[\Psi_{0}]=1\iff\dot{\Psi}_{0}=\dot{\Psi}_{0}^{\text{exact}}
\]
We thus see that the ratio\textbf{ $r_{\mathcal{M}}[\Psi_{0}]$ }is
a convenient measure of the performance of a variational method for
the given dynamical problem.

The above result can be generalized to the case in which the manifold
$\mathcal{M}$ is time-dependent, $\mathcal{M}=\mathcal{M}(t)$, and
the time-derivative of the wavefunction contains both a variational
($\ket{\dot{\Psi}_{v}}\in\text{T}_{0}\mathcal{M}(0)$) and a non-variational
($\ket{\dot{\Psi}_{n}}$) contribution, \emph{i.e.,} $\ket{\dot{\Psi}_{0}}=\ket{\dot{\Psi}_{v}}+\ket{\dot{\Psi}_{n}}$.
In this case energy is not conserved 
\[
\frac{dE_{0}}{dt}=2\Re\braket{\dot{\Psi}_{0}|H|\Psi_{0}}=2\Re\left(\braket{\dot{\Psi}_{n}|(H-i\hbar\partial_{t})\Psi_{0}}\right)
\]
but the error takes yet a simple form
\[
\varepsilon_{\mathcal{M}}^{2}[\Psi_{0}]=\frac{1}{\hbar^{2}}\left(||H\Psi_{0}-i\hbar\dot{\Psi}_{n}||^{2}-\hbar^{2}||\dot{\Psi}_{v}||^{2}\right)
\]
see SM for details. 

\textbf{\emph{Examples}}\emph{. }As a first example, we consider a
simple one-dimensional system whose wavefunction $\ket{\Psi_{0}}$
is constrained to have a Bargmann form\cite{Bargmann1961,Gardiner2004},
$\ket{\Psi_{0}}=C\exp\left(za^{\dagger}\right)\ket{0}$, where the
phonon annhilation operator $a$ reads as $a=\frac{\hat{q}}{2\Delta q}+i\frac{\hat{p}}{2\Delta p}$,
$\hat{q}$ and $\hat{p}$ being the usual coordinate and momentum
operators and $\Delta q$, $\Delta p$ being two parameters satisfying
$\Delta q\Delta p=\hbar/2$, and representing, respectively, the coordinate
and momentum width of the state. Finally, $\ket{0}$ is the vacuum
state ($a\ket{0}=0$) and $C,z\in\mathbb{C}$ parametrize the vector.
This is a semiclassical approximation to the dynamics, also known
as Frozen Gaussian approximation (FGA)\cite{Heller1981}, since the
variational equations of motion reduce to evolution laws for the average
position and momentum of the wavepacket, $q_{0}=2\Delta q\Re z$ and
$p_{0}=2\Delta p\Im z$, respectively. A straightforward calculation
gives the equation of motion for $z$ (see SM for details), $\dot{z}=i\hbar^{-1}\braket{\Psi_{0}|[H,a]|\Psi_{0}}/\braket{\Psi_{0}|\Psi_{0}}$,
and the time-derivative of the wavefunction in the standard \emph{gauge},
$\ket{\dot{\Psi}_{0}^{+}}=\dot{z}(a^{\dagger}-z^{*})\ket{\Psi_{0}}$.
Thus, the error due to the FGA to the dynamics follows as $\hbar^{2}\varepsilon^{2}=\Delta E_{0}^{2}-\hbar^{2}|\dot{z}|^{2}$,
where (for $H=\frac{p^{2}}{2m}+V$) the second term on the r.h.s.
is just the variance of the following variational Hamiltonian
\[
H_{v}=\frac{p_{0}}{m}\delta\hat{p}+\braket{V'}\delta\hat{q}
\]
where $\delta\hat{p}=\hat{p}-p_{0}$, $\delta\hat{q}=\hat{q}-q_{0}$
and $V'=\frac{dV(\hat{q})}{dq}$. The error is easily seen to vanish
when $H$ takes a harmonic form, \emph{i.e.}, $H=H_{\text{HO}}=\hbar\omega a^{\dagger}a+\lambda a^{\dagger}+\lambda^{*}a$
($\omega\in\mathbb{R}$, $\lambda\in\mathbb{C}$), and in general
it reads as, to lowest order in $\Delta q$, 
\[
\hbar\varepsilon\approx\Delta q^{2}\sqrt{\frac{m^{2}\Delta^{4}}{2}+\left(\frac{|V_{0}^{(3)}|^{2}}{6}+\frac{m\Delta^{2}}{2}V_{0}^{(4)}\right)\Delta q^{2}}
\]
where $V_{0}^{(n)}$ is the $n^{\text{th}}$ derivative of the potential
in $q_{0}$, $m\Delta^{2}=V_{0}^{(2)}-m\omega^{2}$, and $\omega=\hbar/2m\Delta q^{2}$
(see SM). In locally harmonic potentials ($V^{(2)}>0$), one may
set $\Delta q$ to make the first term on the r.h.s. vanishing and
obtain $\hbar\varepsilon\approx\hbar^{3}|V_{0}^{(3)}|/8\sqrt{6}[mV_{0}^{(2)}]^{3/2}$
, although this condition only holds at $t=0$ if $\Delta q$ is kept
frozen. 

As a second example, let us consider the general $N-$particle Hamiltonian
$H=\sum_{i=1}^{N}h_{i}+V$ (where $h_{i}$ are one-particle operators
and $V$ is a many-body interaction potential) and the mean-field
\emph{ansatz} of the time-dependent Hartree method, $\ket{\Psi_{0}}=\Pi_{i=1}^{N}\ket{\phi_{i}}$
where the $\phi_{i}$'s are variational single-particle functions
(spf's), subject only to the normalization condition $\braket{\phi_{i}|\phi_{i}}=1$.
Application of the DF condition, Eq. \ref{eq:imaginary DF condition},
gives the equations of motion of the spf's in the form (SM)
\[
i\hbar\ket{\dot{\phi}_{i}}=\left(H_{i}+g_{i}-\bar{E}_{0}\right)\ket{\phi_{i}}
\]
where $H_{i}=\braket{\Psi^{i}|H|\Psi^{i}}$ is the mean-field Hamiltonian
for the $i^{\text{th}}$ degree of freedom ($\ket{\Psi^{i}}=\Pi_{j\neq i}\ket{\phi_{j}}$
is the $i^{\text{th}}$ single-hole wavefunction) and $g_{i}=i\hbar\braket{\phi_{i}|\dot{\phi}_{i}}\in\mathbb{R}$
are arbitrary \emph{gauge} terms that enforce the normalization conditions.
As shown in SM, the total time-derivative of the state vector in the
standard \emph{gauge} follows as
\[
i\hbar\ket{\dot{\Psi}_{0}^{+}}=H_{\text{mf}}^{0}\ket{\Psi_{0}}\ \ \ H_{\text{mf}}^{0}=\sum_{i=1}^{N}\left(H_{i}-\bar{E}_{0}\right),\ \ \ \braket{H_{\text{mf}}}=0
\]
(here $H_{\text{mf}}^{0}$ is the appropriate variational Hamiltonian
for the problem) and thus it holds $\hbar^{2}||\dot{\Psi}_{0}^{+}||^{2}=\Delta E_{\text{mf},0}^{2}$,
where $\Delta E_{\text{mf},0}^{2}=\sum_{i=1}^{N}\Delta E_{i,0}^{2}\ $
and $\Delta E_{i,0}^{2}=\braket{\left(H_{i}-\bar{E}_{0}\right)^{2}}_{0}$
are the one-particle energy fluctuations. Furthermore, since $H-\bar{E}_{0}=H_{\text{mf}}+\Delta V$,
where $\Delta V=V+(N-1)\braket{V}-\sum_{i=1}^{N}v_{i}$ is the zero-mean
fluctuating potential, the energy variance can be given in a simple
form (here $\Delta V_{0}^{2}=\braket{\Delta V^{2}}_{0}$)
\[
\Delta E_{0}^{2}=\Delta E_{\text{mf},0}^{2}+\Delta V_{0}^{2}+2\sum_{i=1}^{N}\Re\braket{H_{i}\Delta V}_{0}
\]
and $r_{\text{mf}}^{2}[\Psi_{0}]\equiv\Delta E_{\text{mf},0}^{2}/\Delta E_{0}^{2}$.
The above expression clearly shows the key role played by the potential
energy fluctuations in limiting the reliability of the mean-field
approach and indicates that 
\[
\varepsilon_{\text{mf}}=\hbar^{-1}\left[\Delta V_{0}^{2}+2\sum_{i=1}^{N}\Re\braket{H_{i}\Delta V}_{0}\right]^{1/2}
\]
is the appropriate expression for the correlation error intrinsic
in the TDH method. Notice that from the inequality $\sum_{i=1}^{N}\Re\braket{H_{i}\Delta V}_{0}=\Re\braket{(H_{\text{mf}}\Delta V}_{0}\leq\Delta E_{\text{mf},0}\Delta V_{0}$
follows a simple lower bound for the $r-$index, namely $r_{\text{mf}}[\Psi_{0}]\geq\Delta E_{\text{mf},0}/(\Delta E_{\text{mf},0}+\Delta V_{0})$. 

Finally, as a last example we consider the error intrinsic to the
adiabatic (Born-Oppenheimer) dynamics, a common strategy to tackle
molecular problems where the electronic degrees of freedom are averaged
out with the well-known \emph{ansatz}
\[
\ket{\Psi_{0}}=\int d\mathbf{X}\psi(\mathbf{X})\ket{\Phi_{n}(\mathbf{X})}\ket{\mathbf{X}}
\]
Here $\mathbf{X}$ represents the nuclear degrees of freedom, and
$\ket{\Phi_{n}(\mathbf{X})}$ is the $n^{\text{th}}$ eigenstate of
the electronic Hamiltonian with clamped nuclei at $\mathbf{X}$, \emph{i.e.},
the electronic operator $h_{\text{el}}(\mathbf{X}$) defined by $\braket{\mathbf{X}|H-T|\mathbf{X'}}=h_{\text{el}}(\mathbf{X})\delta(\mathbf{X}-\mathbf{X'})$,
$H$ being the total Hamiltonian and $T$ the kinetic energy of the
nuclei. Application of the variational principle gives the equation
of motion for the ``nuclear wavefunction'' $\psi(\mathbf{X})$ in
the $n^{\text{th}}$ electronic state
\[
H_{n}\psi=i\hbar\frac{\partial\psi}{\partial t},\ \ \ \ \ H_{n}=\braket{T}_{n}+E_{n}(\mathbf{X})
\]
where $E_{n}(\mathbf{X})$ is the electronic energy and 
\begin{align*}
\braket{T}_{n}= & T-i\sum_{i}\frac{\hbar}{M_{i}}\left<\Phi_{n}|\frac{\partial\Phi_{n}}{\partial R_{i}}\right>_{\text{el}}P_{i}+\\
 & -\sum_{i}\frac{\hbar^{2}}{2M_{i}}\left<\Phi_{n}|\frac{\partial^{2}\Phi_{n}}{\partial R_{i}^{2}}\right>_{\text{el}}
\end{align*}
is a self-adjoint operator, the nuclear kinetic energy operator averaged
over the electronic state\footnote{Here, for the second term on the r.h.s. it holds $\left<\Phi_{n}|\frac{\partial\Phi_{n}}{\partial R_{i}}\right>_{\text{el}}\equiv\Im\left<\Phi_{n}|\frac{\partial\Phi_{n}}{\partial R_{i}}\right>_{\text{el}}$
because of norm conservation. This term vanishes in the presence of
time-reversal invariance (\emph{i.e.}, when the electronic wavefunctions
can be chosen globally real).}. This gives the rate of variation of the wavefunction in the standard
\emph{gauge }as
\[
\hbar^{2}||\dot{\Psi}_{0}^{+}||^{2}=\int d\mathbf{X}\psi^{*}(\mathbf{X})\left[\braket{T}_{n}+(E_{n}(\mathbf{R})-\bar{E}_{0})\right]^{2}\psi(\mathbf{X})
\]
while the energy variance reads as
\[
\Delta E_{0}^{2}=\int d\mathbf{X}\psi^{*}(\mathbf{X})\braket{\left[T+(h_{\text{el}}(\mathbf{X})-\bar{E}_{0})\right]^{2}}_{n}\psi(\mathbf{X})
\]
Hence, the local-in-time error in the adiabatic approximation takes
the form of a nuclear kinetic energy fluctuation term
\[
\varepsilon^{2}[\Psi_{0}]=\frac{1}{\hbar^{2}}\int d\mathbf{X}\psi^{*}(\mathbf{X})\left[\braket{T^{2}}_{n}-\braket{T}_{n}^{2}\right]\psi(\mathbf{X})
\]
This can also be put in a form that makes explicit the contributions
of electronic transitions, that is, upon introducing $\phi_{m\leftarrow n}(\mathbf{X})=\braket{\Phi_{m}|T|\Phi_{n}}_{\text{el}}\psi(\mathbf{X})$,
\[
\varepsilon^{2}[\Psi_{0}]=\frac{1}{\hbar^{2}}\sum_{m\neq n}\int d\mathbf{X}|\phi_{m\leftarrow n}(\mathbf{X})|^{2}
\]
Here, the amplitudes read explicitly as
\begin{align*}
\phi_{m\leftarrow n}(\mathbf{X}) & =-\sum_{i,\alpha}\frac{\hbar^{2}}{2M_{i}}\left[\frac{F_{mn}^{i,\alpha}(\mathbf{X})}{\Delta E_{mn}(\mathbf{X})}\frac{\partial\psi(\mathbf{X})}{\partial\mathbf{X}_{i,\alpha}}\right.\\
 & \left.+B_{mn}^{i,\alpha}(\mathbf{X})\psi(\mathbf{X})\right]
\end{align*}
where $\Delta E_{mn}=E_{m}-E_{n}$, $i$ and $\alpha$ label the nuclei
and their coordinates, respectively, $F_{mn}^{i,\alpha}=\braket{\Phi_{m}|F^{i,\alpha}|\Phi_{n}}$
where $F^{i,\alpha}$ is the operator for the $\alpha$ component
of the force acting on the nucleus $i$, and $B_{mn}^{i,\alpha}=\left<\Phi_{m}|\frac{\partial^{2}\Phi_{n}}{\partial\mathbf{X}_{i,\alpha}^{2}}\right>$. 

\textbf{\emph{Adaptive propagation scheme}}\textbf{s.} Eq. \ref{eq:DF error nice}
represents a rigorous criterion to optimize \emph{on-the-fly} the
computational cost of a quantum dynamical simulation, as it can be
used to re-size the underlying variational manifold in order to keep
the error below a specified ``tolerable'' value (see also Eq. \ref{eq:a posteriori error bound}).
We sketch here its application to a rather popular and quite efficient
variational method for high-dimensional systems, the multiconfiguration
time-dependent Hartree (MCTDH) method \cite{mey90:73,bec00:1,mey09:book,mey12:351}.
In this method the wavefunction takes the form $\ket{\Psi_{0}}=\sum_{I}C_{I}\ket{\Phi_{I}}$
where $C_{I}$'s are complex coefficients, $I=(i_{1},i_{2},..i_{N})$
is a multi-index and $\ket{\Phi_{I}}=\ket{\phi_{i_{1}}\phi_{i_{2}}..\phi_{i_{N}}}$
(where $i_{k}=1,..n_{K}$) are configurations of fully flexible spf's.
Of interest here is the possibility of changing \emph{on-the-fly}
the number of spfs, which means varying both the size of the secular
problem for the amplitude coefficients and the number of spfs to be
optimized. Notice that this would solve from the outset the problem
of regularizing solutions that contain configurations with vanishing
weight. We focus on the ``spawning'' process\cite{men17:113}, \emph{i.e.}
the generation of new spfs and related configurations, which becomes
necessary when, in the course of the dynamics, the local error $\varepsilon$
exceeds some given threshold, thereby signaling the need for a more
flexible manifold. If the main correction comes from single excitations
of the ``occupied'' configurations $\ket{\Phi_{I}}$, the ``best''
spf $\ket{\eta}$ to add to the $k^{\text{th}}$ degree of freedom
is the one the maximizes the expectation value of a certain reduced,
self-adjoint ``rate'' operator $\Gamma^{(k)}$ for the $k^{\text{th}}$
mode (see SM), among those single-particle states that lie in the
orthogonal complement of both the occupied spfs for the $k^{\text{th}}$
mode ($\ket{\phi_{i_{k}}}$, $i_{k}=1,n_{k}$) and their time-derivatives.
The reduced operator reads as 
\[
\Gamma^{(k)}=\sum_{I(k)}\braket{\Phi_{I(k)}|H|\Psi_{0}}\braket{\Psi_{0}|H|\Phi_{I(k)}}
\]
where $\Phi_{I(k)}$ is a $k^{\text{th}}$ hole configuration and
the scalar products are taken over all modes except the $k^{\text{th}}$.
Then, the reduction of the local-in-time (squared) error when adding
such spf is given by $\braket{\eta|\Gamma^{(k)}|\eta}/\hbar^{2}$
(see SM for details). 

\textbf{\emph{Conclusions}}\textbf{. }Variational solutions of the
time-dependent Schr\"{o}dinger equation have an intrinsic measure
of their reliability, a local-in-time error that measures the departure
from the instantaneous exact solution. Simple expressions have been
provided for this error in some relevant cases, with the aim of showing
how the error helps to assess quantitatively the reliability of the
variational method for a given dynamical problem. Future applications
involve the development of adaptive propagation schemes that re\textendash size
\emph{on-the-fly} the variational manifold, and optimize the computational
cost for a target accuracy. 

\bibliographystyle{apsrev4-1}
%

\section*{Supplemental Material}

\subsection*{\emph{A posteriori error bound}}

Following Ref. \cite{LubichBook}, let $\ket{\Psi(t)}$ and $\ket{\tilde{\Psi}(t)}$
be, respectively, an approximate and the exact solution of the TDSE
with the same initial state, $\ket{\Psi(0)}=\ket{\tilde{\Psi}(0)}\equiv\ket{\Psi_{0}}$
and $\ket{\Delta\Psi}=\ket{\Psi(t)}-\ket{\tilde{\Psi}(t)}$. From
the identity 
\[
i\hbar\ket{\Delta\dot{\Psi}}-H\ket{\Delta\Psi}=i\hbar\ket{\dot{\Psi}}-H\ket{\Psi}
\]
it follows
\[
\hbar\Re\braket{\Delta\Psi|\Delta\dot{\Psi}}=\Im\braket{\Delta\Psi|i\hbar\dot{\Psi}-H\Psi}
\]
Here, $\Re\braket{\Delta\Psi|\Delta\dot{\Psi}}=\frac{1}{2}\frac{d}{dt}||\Delta\Psi||^{2}=||\Delta\Psi||\frac{d||\Delta\Psi||}{dt}$
and thus
\[
\hbar||\Delta\Psi||\frac{d||\Delta\Psi||}{dt}=\Im\braket{\Delta\Psi|i\hbar\dot{\Psi}-H\Psi}\leq||\Delta\Psi||\ ||i\hbar\dot{\Psi}-H\Psi||
\]
\emph{i.e.}, 
\[
\frac{d}{dt}||\Delta\Psi||\leq\frac{1}{\hbar}||i\hbar\dot{\Psi}-H\Psi||
\]
which integrated gives 
\[
||\Delta\Psi||\leq\frac{1}{\hbar}\int_{0}^{t}||i\hbar\dot{\Psi}(\tau)-H\Psi(\tau)||d\tau
\]
When $\ket{\Psi(t)}\in\mathcal{M}$ is a variational solution the
integrand on the r.h.s. takes at any time its minimum value and it
is just the local-in-time error $\varepsilon_{\mathcal{M}}[\Psi(t)]$
defined in the main text, Eq. \ref{eq:DF error}. 

The above bound also contraints the error in autocorrelation functions
(here and below $||\Psi_{0}||=1$)
\[
|\braket{\Psi(t)|\Psi_{0}}-\braket{\tilde{\Psi}(t)|\Psi_{0}}|=|\braket{\Delta\Psi|\Psi_{0}}|\leq||\Delta\Psi||
\]
and in the average values of any bounded observable, 
\begin{align*}
|\braket{\Psi(t)|A|\Psi(t)}-\braket{\tilde{\Psi}(t)|A|\tilde{\Psi}(t)}|=\\
=|\braket{\Delta\Psi|A|\Psi(t)}+\braket{\tilde{\Psi}(t)|A|\Delta\Psi}|\\
\leq||\Delta\Psi||\left(||A\Psi||+||A\tilde{\Psi}||\right)\\
\leq2||A||_{\infty}||\Delta\Psi||
\end{align*}
where $||A||_{\infty}$ is the operator norm. 

\subsection*{\emph{Error and energy drift with time-dependent manifolds} }

We address here in some detail the situation where the manifold $\mathcal{M}$
is time-dependent and the time-derivative of the wavefunction contains
both a variational and a non-variational contribution
\[
\ket{\dot{\Psi}_{0}}=\ket{\dot{\Psi}_{0}^{T}}+\ket{\dot{\Phi}_{0}}
\]
(here the superscript \emph{T} reminds us that $\ket{\dot{\Psi}_{0}^{T}}\in\text{T}_{0}\mathcal{M}$,
the space tangent to $\mathcal{M}(t)$ at $t=0$) . This may happen,
for instance, when the manifold is described by a set of variational
parameters $x_{1},x_{2},..x_{N}$ and a number of additional time-dependent
parameters $y_{1},y_{1},..y_{M}$ which, for computational efficiency,
are evolved according to some physically sound law (``guided'' parameters),
simpler than the variational equations of motion. In such circumstances,
the (partial) variational condition $\Im\braket{\dot{\Psi}_{0}^{T}|(i\hbar\partial_{t}-H)\Psi_{0}}=0$
leads to
\[
\hbar\braket{\dot{\Psi}_{0}^{T}|\dot{\Psi}_{0}^{T}}=\Im\braket{\dot{\Psi}_{0}^{T}|H\Psi_{0}-i\hbar\dot{\Phi}_{0}}
\]
which generalizes Eq. \ref{eq:novel condition}. Hence, for the error
it follows 
\begin{align*}
\braket{\left(i\hbar\dot{\Phi}_{0}-H\Psi_{0}\right)+i\hbar\dot{\Psi}_{0}^{T}|\left(i\hbar\dot{\Phi}_{0}-H\Psi_{0}\right)+i\hbar\dot{\Psi}_{0}^{T}} & =\\
||i\hbar\dot{\Phi}_{0}-H\Psi_{0}||^{2}+\hbar^{2}||\dot{\Psi}_{0}^{T}||^{2}-2\hbar\Im\braket{\dot{\Psi}_{0}^{T}|i\hbar\dot{\Phi}_{0}-H\Psi_{0}}
\end{align*}
and thus
\[
\varepsilon_{\mathcal{M}}^{2}[\Psi_{0}]=\frac{1}{\hbar^{2}}\left(||H\Psi_{0}-i\hbar\dot{\Phi}_{0}||^{2}-\hbar^{2}||\dot{\Psi}_{0}^{T}||^{2}\right)
\]
(cfr. Eq. \ref{eq:DF error}) and the inequality
\[
\hbar||\dot{\Psi}_{0}^{T}||\leq||H\Psi_{0}-i\hbar\dot{\Phi}_{0}||
\]
that can be considered a generalization of the boundedness theorem
above to the case in which the manifold is time-dependent. Here, the
appearance of $i\hbar\dot{\Phi}_{0}$ on the r.h.s. of the inequality
can be understood in the limiting case where the non-variational time-derivative
comes from an effective Hamiltonian, \emph{i.e.} $i\hbar\ket{\dot{\Phi}_{0}}=H^{\text{eff}}\ket{\Psi_{0}}$,
since in such case the above inequality reduces to 
\[
\hbar||\dot{\Psi}_{0}^{T}||\leq||(H-H^{\text{eff}})\Psi_{0}||
\]
a rather reasonable result. 

It is instructive at this point to consider these results in view
of the energy conservation since when the wavefunction contains ``guided''
parameters energy is no longer conserved. Thus in the following we
assume that three variational principles are equivalent to each other
on $\mathcal{M}$ and consider the energy change per unit time 
\begin{align*}
W_{0} & =\frac{dE_{0}}{dt}=2\Re\braket{\dot{\Psi}_{0}|H|\Psi_{0}}\\
 & =2\Re\braket{\dot{\Psi}_{0}^{T}|H|\Psi_{0}}+2\Re\braket{\dot{\Phi}_{0}|H|\Psi_{0}}\\
 & =2\Re\left(\braket{\dot{\Phi}_{0}|(H-i\hbar\partial_{t})\Psi_{0}}\right)
\end{align*}
where the last equality follow from the Dirac-Frenkel condition $\braket{\dot{\Psi}_{0}^{T}|(i\hbar\partial_{t}-H)\Psi_{0}}=0$,
namely from 
\[
\braket{\dot{\Psi}_{0}^{T}|H|\Psi_{0}}=i\hbar\braket{\dot{\Psi}_{0}^{T}|\dot{\Psi}_{0}}=i\hbar\braket{\dot{\Psi}_{0}|\dot{\Psi}_{0}}-i\hbar\braket{\dot{\Phi}_{0}|\dot{\Psi}_{0}}
\]
When optimizing also w.r.t. $\ket{\delta\dot{\Phi}_{0}}$, the above
equation shows that the (magnitude of the) energy drift is stationary
at the variational minimum 
\[
\delta W_{0}=-2\Re\left(\braket{\delta\dot{\Phi}_{0}|(i\hbar\partial_{t}-H)\Psi_{0}}\right)=0
\]
a trivial result because we already known that $|W_{0}|$ is actually
at its minimum under such circumstances ( $|W_{0}|=0$ ), but, in
general, it shows that\textbf{ }optimizing the guide (under given
constraints) minimizes the energy dritft.\textbf{ }In this context
it is worth noticing that for a variational solution it must hold
\[
|W_{0}|\leq2||\dot{\Phi}_{0}||\ ||(i\hbar\partial_{t}-H)\Psi_{0}||=2\hbar\varepsilon_{\mathcal{M}}[\Psi_{0}]||\dot{\Phi}_{0}||
\]
that can be converted into a lower bound on the variational solution
in terms of energy drift, 
\[
\varepsilon_{\mathcal{M}}[\Psi_{0}]\geq\frac{|W_{0}|}{2\hbar||\dot{\Phi}_{0}||}
\]
Thus, optimization of the guide (minimization of $|W_{0}|$) effectively
lowers the bound by reducing the error contribution due to the non-conservation
of the energy. 

\subsection*{\emph{Mean-field approximation}}

Let us consider the general $N-$particle Hamiltonian $H=\sum_{i=1}^{N}h_{i}+V$,
where $h_{i}$ are one-particle operators and $V$ is a many-body
interaction potential, and the mean-field \emph{ansatz} of the time-dependent
Hartree method,
\[
\ket{\Psi_{0}}=\Pi_{i=1}^{N}\ket{\phi_{i}}
\]
where $\phi_{i}'$s are variational single-particle functions, subjected
only to the normalization condition $\braket{\phi_{i}|\phi_{i}}=1$
that is enforced through the \emph{guage} terms $i\hbar\braket{\phi_{i}|\dot{\phi}_{i}}=g_{i}\in\mathbb{R}$.
Application of the DF condition, Eq. \ref{eq:imaginary DF condition},
gives the equations of motion of the spf's. To this end, it is worth
noticing that it suffices to consider only the special (complex) spf's
variations satisfying $\braket{\delta\phi_{i}|\phi_{i}}=0$\emph{
}(\emph{i.e.} $\ket{\delta\phi_{i}}\in V_{i}'\equiv\{\ket{\phi_{i}}\}^{\perp}$)
along with the Dirac-Frenkel condition (Eq. \ref{eq:imaginary DF condition})
since the general stationary condition adds nothing (this is evident
upon introducing the projector $P_{i}=\ket{\phi_{i}}\bra{\phi_{i}}$
and noticing that $\Re\braket{\delta_{i}\Psi_{0}|P_{i}\left(i\hbar\partial_{t}-H\right)|\Psi_{0}}\equiv0$
when $\Re\braket{\delta\phi_{i}|\phi_{i}}=0$ and $i\hbar\braket{\phi_{k}|\dot{\phi}_{k}}=g_{k}\in\mathbb{R}$). 

Thus, the requirement $\left(i\hbar\ket{\dot{\phi}_{i}}-H_{i}\ket{\phi_{i}}\right)\in V_{i}^{\perp}=\{\ket{\phi_{i}}\}^{\perp\perp}$
gives $i\hbar\ket{\dot{\phi}_{i}}-H_{i}\ket{\phi_{i}}=\alpha\ket{\phi_{i}}$,
where $\alpha$ is easily found to be $\alpha=i\hbar\braket{\phi_{i}|\dot{\phi}_{i}}-\bar{E}_{0}\equiv g_{i}-\bar{E}_{0}$,
and the equations of motion take the form
\[
i\hbar\ket{\dot{\phi}_{i}}=\left(H_{i}+g_{i}-\bar{E}_{0}\right)\ket{\phi_{i}}
\]
where $H_{i}=\braket{\Psi^{i}|H|\Psi^{i}}$ is the mean-field Hamiltonian
for the $i$th degree of freedom ($\ket{\Psi^{i}}=\Pi_{j\neq i}\ket{\phi_{j}}$
is the $i$th single-hole wavefunction) and $E=\braket{\Psi|H|\Psi}\equiv\braket{H_{i}}$.
It follows that the total time-derivative of the state vector satisfies
\[
i\hbar\ket{\dot{\Psi}}=H_{\text{mf}}\ket{\Psi}
\]
where the mean-field (total) Hamiltonian $H_{\text{mf}}$ reads as
\[
H_{\text{mf}}=\sum_{i=1}^{N}(H_{i}+g_{i}-\bar{E}_{0})
\]
The optimal \emph{gauge} condition on the total wavefunction requires
$\sum_{i}g_{i}=0$ and thus, introducing now the initial time $t=0$,
\[
i\hbar\ket{\dot{\Psi_{0}^{+}}}=H_{\text{mf}}^{0}\ket{\Psi_{0}}\ \ \ H_{\text{mf}}^{0}=\sum_{i=1}^{N}\left(H_{i}-\bar{E}_{0}\right),\ \ \ \braket{H_{\text{mf}}}=0
\]
Now, the mean-field Hamiltonians $H_{i}$ read as $H_{i}=h_{i}+\sum_{j}\epsilon_{j}-\epsilon_{i}+v_{i}$
(where $\epsilon_{i}=\braket{\phi_{i}|h_{i}|\phi_{i}}$ is the average
one-particle energy on the $i$th degree and $v_{i}=\braket{\Psi^{i}|V|\Psi^{i}}$
is the $i$th mean-field potential) hence it is easy to check that
it holds $H-\bar{E}_{0}=H_{\text{mf}}+\Delta V$ where 
\[
\Delta V=V+(N-1)\bar{V}-\sum_{i=1}^{N}v_{i}
\]
is the zero-mean fluctuating potential\textbf{ }($\bar{V}=\braket{V}\equiv\braket{v_{i}}$
for any $i$). Thus,
\[
\Delta E_{0}^{2}=\Delta E_{\text{mf},0}^{2}+\braket{\Delta V^{2}}_{0}+2\sum_{i=1}^{N}\Re\braket{H_{i}\Delta V}_{0}
\]
and 
\[
r_{\text{mf}}^{2}[\Psi_{0}]=\frac{\Delta E_{\text{mf},0}^{2}}{\Delta E_{\text{mf},0}^{2}+\Delta V_{0}^{2}+2\sum_{i=1}^{N}\Re\braket{H_{i}\Delta V}_{0}}
\]
where $\Delta V_{0}^{2}=\braket{\Delta V^{2}}_{0}$ and $\Delta E_{\text{mf},0}^{2}\equiv\sum_{i=1}^{N}\Delta E_{i,0}^{2}\ \ $,
being $\Delta E_{i,0}^{2}=\braket{\left(H_{i}-\bar{E}_{0}\right)^{2}}_{0}$
the one-particle energy fluctuations (one may further notice that
they consist of both a ``kinetic'' and a ``potential'' term, since
$H_{i}-E=\left(h_{i}-\epsilon_{i}\right)+\left(v_{i}-\bar{V}\right)$). 

\subsection*{\emph{Coherent state (or Frozen Gaussian) approximation}}

We detail here the case of a coherent state approximation to the dynamics
by considering a situation slightly more general than the one presented
in the main text, namely a system with two degrees of freedom to which
we apply the mean-field approximation 
\[
\ket{\Psi_{0}}=\ket{\phi_{1}}\ket{\phi_{2}}
\]
and force the single particle function of the second degree to take
the form of a normalized coherent-state (CS) 
\[
\ket{\phi_{2}}\equiv\ket{\theta,z}=\exp\left(i\theta-\frac{|z|^{2}}{2}+za^{\dagger}\right)\ket{0}
\]
This wavefunction is a ``precusor'' of the Ehrenfest method, with
$\ket{\phi_{1}}$ describing an ``electronic'' system and $\ket{\phi_{2}}$
a ``semiclassical'' nuclear degree of freedom. The equation of motion
for $\ket{\phi_{2}}$ (or, better, $z$) can be derived either from
the Dirac-Frenkel condition, Eq. \ref{eq:DF condition}, or from the
McLachlan minimum-distance condition, Eq. \ref{eq:McLaclhan condition}.
For illustrative purposes we follow the second route, and consider
\[
\bra{\phi_{1}\delta\dot{\phi}_{2}}\left[i\hbar\ket{\dot{\phi}_{1}\phi_{2}}+i\hbar\ket{\phi_{1}\dot{\phi}_{2}}-H\ket{\phi_{1}\phi_{2}}\right]=0
\]
which is the appropriate condition for optimizing $\ket{\phi_{2}}$.
Notice that, though not evident from the chosen parametrization, the
CS variations can be considered complex, as seen by considering the
unnormalized Bargmann vectors $C\exp(za^{\dagger})\ket{0}$ and the
complex analytic parametrization $(C,z)\in\mathbb{C}^{2}$ (as mentioned
in the main text). Some lenghty but simple algebra leads to 
\begin{align*}
-i\delta\dot{\theta}\left\{ -\hbar\dot{\theta}-\hbar\Im\left(\dot{z}z^{*}\right)+g-\bar{E}_{0}\right\} +\\
\delta\dot{z}^{*}\left\{ i\hbar\dot{z}+\frac{z}{2}\left[-\hbar\dot{\theta}-\hbar\Im\left(\dot{z}z^{*}\right)+g+\bar{E}_{0}\right]-\braket{\phi_{2}|aH_{\text{cl}}|\phi_{2}}\right\} +\\
\delta\dot{z}\left\{ \frac{z^{*}}{2}\left[\hbar\dot{\theta}+\hbar\Im\left(\dot{z}z^{*}\right)-g+\bar{E}_{0}\right]\right\} =0
\end{align*}
where $g=i\hbar\braket{\phi_{1}|\dot{\phi}_{1}}$, $H_{\text{cl}}=\braket{\phi_{1}|H|\phi_{1}}$
and $\bar{E}_{0}=\braket{\Psi_{0}|H|\Psi_{0}}$. Hence, the optimal
gauge $\theta$ (a concept that only becomes meaningful in view of
computing an error) is such that 
\[
\hbar\dot{\theta}=-\hbar\Im\left(\dot{z}z^{*}\right)+g-\bar{E}_{0}
\]
and the stationary condition reduces to 
\[
i\hbar\dot{z}+\bar{E}_{0}-\braket{\phi_{2}|aH_{\text{cl}}|\phi_{2}}=0
\]
\emph{i.e.}, 
\[
\dot{z}=\frac{i}{\hbar}\braket{\phi_{2}|[H_{\text{cl}},a]|\phi_{2}}
\]
It follows
\begin{align*}
i\hbar\ket{\dot{\Psi}_{0}} & =\\
\left[(H_{q}+g-\bar{E}_{0})+(-\hbar\dot{\theta}-i\hbar\Re\left(\dot{z}z^{*}\right)+i\hbar\dot{z}a^{\dagger})\right]\ket{\Psi_{0}}
\end{align*}
(where $H_{q}=\braket{\phi_{2}|H|\phi_{2}}$) and thus, upon replacing
$\dot{\theta}$ with its optimal value, 
\[
i\hbar\ket{\dot{\Psi}_{0}}\equiv\left[H_{q}+i\hbar\dot{z}(a^{\dagger}-z^{*})\right]\ket{\Psi_{0}}
\]
in such a way that it holds $i\hbar\braket{\Psi_{0}|\dot{\Psi}_{0}}=\bar{E}_{0}$
as required by the minimum-distance principle. Note that the \emph{gauge}
term $\theta$ is irrelevant for the parameter dynamics, and can be
safely neglected when deriving the equation of motion for $z$ from
the Dirac-Frenkel condition. However, such a term is needed in order
to make $\ket{\dot{\Psi}_{0}}$ appropriate for computing the error,
and needs to be obtained separately when using the Dirac-Frenkel principle. 

Finally, with the Hamiltonian referenced to $E$, we write the variational
error using Eq. \ref{eq:DF error nice} where now
\[
i\hbar\ket{\dot{\Psi}_{0}^{+}}=\left[H_{q}-\bar{E}_{0}+i\hbar\dot{z}(a^{\dagger}-z^{*})\right]\ket{\Psi_{0}}
\]
leads to the simple expression 
\[
\hbar^{2}||\Psi_{0}^{+}||^{2}=\braket{\Psi_{0}|\left(H_{q}-\bar{E}_{0}\right)^{2}|\Psi_{0}}+\hbar^{2}|\dot{z}|^{2}
\]
By comparing this expression with the above one obtained for the general
TDH case one finds that 
\[
\varepsilon_{z}^{2}=\braket{\Psi_{0}|\left(\frac{H_{\text{cl}}-\bar{E}_{0}}{\hbar}\right)^{2}|\Psi_{0}}-|\dot{z}|^{2}\geq0
\]
is the genuine error due to the coherent-state approximation. The
case considered in the main text can be obtained by setting $\ket{\phi_{1}}\equiv1$,
$g=0$, $H_{q}=\bar{E}_{0}$ and $H_{\text{cl}}=H$. 

\subsection*{\emph{Local-in-time error in the FGA}}

The local-in-time error derived above, $\hbar^{2}\varepsilon^{2}=\Delta E_{0}^{2}-\hbar^{2}|\dot{z}|^{2}$,
is easily seen to vanish when the Hamiltonian takes a harmonic oscillator
(HO) form, $H=H_{\text{HO}}=\hbar\omega a^{\dagger}a+2\hbar\Re(\lambda a^{\dagger})$.
This rather general result in this context follows easily by observing
that, on the one hand, it holds
\[
\dot{z}_{\text{HO}}=-i\left(\omega a+\lambda\right)
\]
and, on the other hand, 
\[
\left(H_{\text{HO}}-E_{\text{HO}}\right)\ket{z}=\hbar\left(\omega z+\lambda\right)\left(a^{\dagger}-z^{*}\right)\ket{z}
\]

In view of the above, we write $H=\frac{p^{2}}{2m}+V=H_{\text{HO}}+W$,
where $W$ is assumed to be local, $W=W(\hat{q})$ (see below). We
find, on the one hand, $\dot{z}=\dot{z}_{\text{HO}}-\frac{i}{2\Delta p}\braket{W'}$
and, on the other hand,
\[
\Delta E_{0}^{2}=\Delta E_{\text{HO}}^{2}+\Delta W^{2}+2\Re\braket{z|\left(W-\braket{W}\right)\left(H_{\text{HO}}-E_{\text{HO}}\right)|z}
\]
The last term on the r.h.s. can be rearranged into
\begin{align*}
\Re\braket{z|\left(W-\braket{W}\right)\left(H_{\text{HO}}-E_{\text{HO}}\right)|z} & =\\
\hbar\Re\left[\left(\omega z+\lambda\right)\braket{z|\left(W-\braket{W}\right)a^{\dagger}|z}\right]
\end{align*}
where $\braket{z|\left(W-\braket{W}\right)a^{\dagger}|z}\equiv\frac{\hbar}{2\Delta p}\braket{W'}$.
It follows
\[
\Delta E_{0}^{2}-\hbar^{2}|\dot{z}|^{2}=\Delta W^{2}-\Delta x^{2}\braket{W'}^{2}
\]

Next, we choose $H_{\text{HO}}$ such that $H-H_{\text{HO}}$ is a
purely local potential. To this end we set $\Delta q^{2}=\hbar/2m\omega$
and, for $H_{\text{HO}}$ in the form $H_{\text{HO}}=\frac{p^{2}}{2m}+\frac{m\omega^{2}}{2}(q-\bar{q})^{2}$,
we obtain $W=V-\frac{m\omega^{2}}{2}\left(q-\bar{q}\right)^{2}$ and
then fix $\bar{q}$ by enforcing the condition $\braket{W'}=0$, \emph{i.e.},
\[
\bar{q}=q_{0}-\frac{\braket{V'}}{m\omega^{2}}
\]
where $q_{0}=\braket{z|\hat{q}|z}$. This reduces the problem of finding
the error to that of computing $\Delta W^{2}$ . Upon using the condition
above $m\omega^{2}(q_{0}-\bar{q})=\braket{V'}$, we readily find
\[
W-\braket{W}=V-\braket{V'}\delta\hat{q}-\frac{m\omega^{2}}{2}\left(\delta\hat{q}^{2}-\Delta q^{2}\right)
\]
where $\delta\hat{q}=q-q_{0}$. Finally, expanding the potential around
$q_{0}$, squaring and averaging 
\[
\Delta W^{2}\approx\frac{m^{2}\Delta^{4}}{2}\Delta q^{4}+\left(\frac{|V_{0}^{(3)}|^{2}}{6}+\frac{m\Delta^{2}}{2}V_{0}^{(4)}\right)\Delta q^{6}
\]
where $m\Delta^{2}:=V''(q_{0})-m\omega^{2}$, $V_{0}^{(n)}$ is a
shorthand for the $n^{\text{th}}$ derivative of the potential evaluated
in $q_{0}$ and $\braket{(q-q_{0})^{2n}}=\Delta q^{2n}(n-1)!!$ for
$n=2$ has been used. 

In closing this section, we notice that from the variational equation
of motion it follows
\[
i\hbar\ket{\dot{\Psi}_{0}^{+}}=\braket{[a,H]}(a^{\dagger}-z^{*})\ket{\Psi_{0}}
\]
and thus 
\[
H_{v}=\braket{[a,H]}(a^{\dagger}-z^{*})-\braket{[a^{\dagger},H]}(a-z)
\]
is the appropriate ``variational'' Hamiltonian. Introducing $\delta\hat{q}$
and $\delta\hat{p}=\hat{p}-p_{0}$, and using the expression of $H$
above, one easily finds
\[
H_{v}=\frac{p_{o}}{m}\delta\hat{p}+\braket{V'}\delta\hat{q}
\]
which is a kind of Hamiltonian linearized around the average position
and momentum of the wavepacket.

\subsection*{\emph{Adiabatic approximation}}

A key quantity in the adiabatic approximation is the kinetic energy
operator ``reduced'' with respect to the electronic coordinates,
$\braket{T}_{nm}=\braket{\Phi_{n}(\mathbf{X})|T|\Phi_{m}(\mathbf{X})}$.
Using $i$ to label the nuclear coordinates with mass $M_{i}$ we
obtain
\begin{align*}
\braket{T}_{nm} & =\delta_{nm}T-i\hbar\sum_{i}\frac{1}{M_{i}}\left<\Phi_{n}(\mathbf{X})|\frac{\partial\Phi_{m}}{\partial X_{i}}\right>P_{i}+\\
 & -\hbar^{2}\sum_{i}\frac{1}{2M_{i}}\left<\Phi_{n}(\mathbf{X})|\frac{\partial^{2}\Phi_{m}}{\partial X_{i}^{2}}\right>
\end{align*}
where $P_{i}$ is a nuclear momentum operator. Here, for $m\ne n$
the second term can also be written in a form that makes explicit
the energy differences, since it holds
\[
\left<\Phi_{n}|\left[\frac{\partial}{\partial X_{i}},h_{\text{el}}(\mathbf{X})\right]|\Phi_{m}\right>\equiv\Delta E_{mn}(\mathbf{X})\left<\Phi_{n}|\frac{\partial\Phi_{m}}{\partial X_{i}}\right>
\]
and, on the other hand, $\left[\frac{\partial}{\partial X_{i}},h_{\text{el}}(\mathbf{X})\right]=-F^{i}$
where $F^{i}$ is the one-electron operator representing the force
acting on $\mathbf{X}_{i}$. The operators $\braket{T}_{nm}$ satisfy
\[
\braket{T}_{nm}^{\dagger}=\braket{T}_{mn}
\]
as can be readily checked by either its definition or a direct calculation.
In the latter case, notice that one needs the identities 
\[
\left<\Phi_{n}|\frac{\partial\Phi_{m}}{\partial X_{i}}\right>+\left<\frac{\partial\Phi_{n}}{\partial X_{i}}|\Phi_{m}\right>=0
\]
\[
\left<\Phi_{n}|\frac{\partial^{2}\Phi_{m}}{\partial X_{i}^{2}}\right>+2\left<\frac{\partial\Phi_{n}}{\partial X_{i}}|\frac{\partial\Phi_{m}}{\partial X_{i}}\right>+\left<\frac{\partial^{2}\Phi_{n}}{\partial X_{i}^{2}}|\Phi_{m}\right>=0
\]
that follow from the orthonormality of the electronic states (the
first make also the diagonal term $\left<\Phi_{n}(\mathbf{X})|\frac{\partial\Phi_{n}}{\partial X_{i}}\right>$
vanishing in the presence of time-reversal invariance).

Finally, in the main text, we have used 
\begin{align*}
\braket{T^{2}}_{nn}-\braket{T}_{nn}^{2} & =\sum_{m\neq n}\braket{T}_{nm}\braket{T}_{mn}\\
 & \equiv\sum_{m\neq n}\braket{T}_{mn}^{\dagger}\braket{T}_{mn}
\end{align*}
to rewrite the error in terms of contributing electronic transitions
$||\braket{T}_{mn}\psi||^{2}$.

\subsection*{\emph{Spawning in MCTDH} }

We sketch here a possible ``spawning'' algorithm in propagating
high-dimensional wavepackets of the multiconfiguration time-dependent
Hartree (MCTDH) type using the error expresion provided by Eq. \ref{eq:DF error nice}.
In this method the wavefunction takes the form $\ket{\Psi_{0}}=\sum_{I}C_{I}\ket{\Phi_{I}}$
where $C_{I}$'s are complex coefficients, $I=(i_{1},i_{2},..i_{N})$
is a multi-index and $\ket{\Phi_{I}}=\ket{\phi_{i_{1}}\phi_{i_{2}}..\phi_{i_{N}}}$
(where $i_{k}=1,..n_{K}$) are configurations of fully flexible single-particle
functions. We call $\ket{\phi_{i_{k}}}$ ($i_{k}=1,n_{k}$) the ``occupied''
spfs for the $k^{\text{th}}$ mode, and $\ket{\Phi_{I}}$ the ``occupied''
configurations. The scalar product over the $k^{\text{th}}$ degree
of freedom 
\[
\braket{\phi_{i_{k}}|\Phi_{J}}=\begin{cases}
0 & \text{if }i_{k}\notin J\\
\ket{\Phi_{J_{k}}^{(k)}} & \text{if }i_{k}\in J
\end{cases}
\]
defines the single-hole configuration $\ket{\Phi_{J_{k}}^{(k)}}$
(and the $N-1$ dimensional multi-index $J_{k}=j_{1}..j_{k-1}j_{k+1}..j_{N}$)
with the spf for the $k^{\text{th}}$ mode removed. When ``spawning''
is required (\emph{i.e.}, when the error $\varepsilon$ exceeds some
given threshold) new spfs are introduced and the set of configurations
enlarged, 
\[
\ket{\Psi_{0}}\rightarrow\ket{\Psi'_{0}}=\ket{\Psi_{0}}+\ket{\delta\Psi_{0}}\ \ \ \ \ket{\delta\Psi_{0}}=\sum_{J}D_{J}\ket{\delta\Phi_{J}}
\]
where $\ket{\delta\Phi_{J}}$'s have one or more occupied spfs replaced
by newly generated ones. At the time of spawning, however, such an
addition does \emph{not} modify the wavefunction ($D_{J}\equiv0$)
but only its time-derivative 
\[
\ket{\dot{\Psi}_{0}}\rightarrow\ket{\dot{\Psi}'_{0}}=\ket{\dot{\Psi}_{0}}+\ket{\delta\dot{\Psi}_{0}}\ \ \ \ \ket{\delta\dot{\Psi}_{0}}=\sum_{J}\dot{D}_{J}\ket{\delta\Phi_{J}}
\]
We consider one additional spf per mode at a time, call it $\ket{\eta_{k}}$
for the $k^{\text{th}}$ mode, and assume that the main contribution
comes through single excitations, i.e., 
\[
\ket{\delta\dot{\Psi}_{0}}\approx\sum_{k}\ket{\delta\dot{\Psi}_{0}^{k}}=\sum_{k}\sum_{J}\dot{D}_{J}^{(k)}\ket{\eta_{k}\Phi_{J}^{(k)}}
\]
where $J$ is now a $N-1$ dimensional index and $\ket{\Phi_{J}^{(k)}}$
a $k^{\text{th}}$-single-hole configuration. If $\ket{\eta_{k}}$
is chosen to be orthogonal to both the occupied spfs ($\ket{\phi_{i_{k}}}$,
$i_{k}=1,n_{k}$) and their time-derivative ($\ket{\dot{\phi}_{i_{k}}}$,
$i_{k}=1,n_{k}$) the above time-derivative is orthogonal to 
\[
\ket{\dot{\Psi}_{0}}=\sum_{I}\dot{C}_{I}\ket{\Phi_{I}}+\sum_{k}\sum_{I}C_{I}\ket{\dot{\phi}_{i_{k}}}\ket{\Phi_{I_{k}}^{(k)}}
\]
and thus 
\[
||\dot{\Psi}'_{0}||^{2}=||\dot{\Psi}{}_{0}||^{2}+\sum_{k}||\mathbf{D}^{(k)}||^{2}\ \ \ \text{where }||\mathbf{D}^{(k)}||^{2}=\sum_{J}|\dot{D}_{J}^{(k)}|^{2}
\]
On the other hand, the amplitude coefficients of the newly introduced
configurations follow from the secular problem
\[
i\hbar\dot{D}_{J}^{(k)}=\braket{\eta_{k}\Phi_{J}^{(k)}|H|\Psi_{0}}
\]
hence 
\[
\hbar^{2}||\mathbf{D}^{(k)}||^{2}=\sum_{J}\braket{\eta_{k}\Phi_{J}^{(k)}|H|\Psi_{0}}\braket{\Psi_{0}|H|\eta_{k}\Phi_{J}^{(k)}}
\]
This suggests to introduce a reduced, self-adjoint operator for the
$k^{\text{th}}$ mode 
\[
\Gamma^{(k)}=\sum_{J}\braket{\Phi_{J}^{(k)}|H|\Psi_{0}}\braket{\Psi_{0}|H|\Phi_{J}^{(k)}}
\]
(where the scalar products are now over all modes except the $k^{\text{th}}$
one) in such a way that it holds 
\[
\hbar^{2}||\mathbf{D}^{(k)}||^{2}=\braket{\eta_{k}|\Gamma^{(k)}|\eta_{k}}=\gamma_{k}
\]
Accordingly, the original local-in-time error $\hbar^{2}\varepsilon^{2}=\Delta E_{0}^{2}-||\dot{\Psi}{}_{0}||^{2}$
transforms, upon spawning, into 
\[
\hbar^{2}\varepsilon'^{2}=\hbar^{2}\varepsilon{}^{2}-\sum_{k}\gamma_{k}
\]
(notice that the added spfs do not modify $\ket{\Psi_{0}}$, hence
neither the average energy nor its variance). One can thus maximize
the error reduction by choosing, for each mode, the eigenvectors of
maximum value of the operator $\Gamma^{(k)}$ (in the appropriate
residual space of the $k^{\text{th}}$ mode). 
\end{document}